\documentclass[aps,twocolumn,showpacs]{revtex4}
\usepackage[dvips]{graphicx}

\begin{document}

\title{Strange eigenstates and anomalous transport in a Koch fractal with 
hierarchical interaction}
\author{Arunava Chakrabarti}
\affiliation{
Department of Physics, University of Kalyani, Kalyani,
West Bengal-741 235, India}

\begin{abstract}
\begin{center}
{\bf Abstract}
\end{center}
Stationary states of non-interacting electrons on a Koch fractal
are investigated within a tight binding approach. It is observed that 
if a hierarchically long range hopping is allowed, a suitable correlation 
between the parameters defining the Hamiltonian leads to spectacular 
changes in the transport properties of finite, but arbitrarily large 
fractals. Topologically identical structures, that are found to support the 
same distribution of the amplitudes of eigenstates, are conducting in some cases
and insulating in the others, depending on the choice of the hierarchy parameter.
The values of the hierarchical parameter themselves display a self-similar, fractal character.  
\end{abstract}
\vskip .25in
\noindent
\maketitle
\vskip .3in
\section{Introduction}
The localization 
of single particle states in a disordered system, viz. the Anderson localization~\cite{pwd}.
has been explored extensively since its inception way back in 1958. 
The investigation 
didn't remain confined to 
 the cases of randomly disordered systems
in one, two or three dimensions alone, but also encompassed  
quasi one dimensional systems with topological disorder~\cite{guinea}-\cite{hjort}, 
 quasi-periodic chains~\cite{koh}-\cite{luck}  and deterministic fractals~\cite{domany}-
\cite{schwalm3}.   
The basic result is that, all the single particle eigenstates of one and 
two dimensional randomly 
disordered lattices should be exponentially localized irrespective of the strength of 
disorder, while in three dimensions there is a metal-insulator transition.  
\begin{figure}[ht]
{\centering \resizebox*{8cm}{7.5cm}{\includegraphics{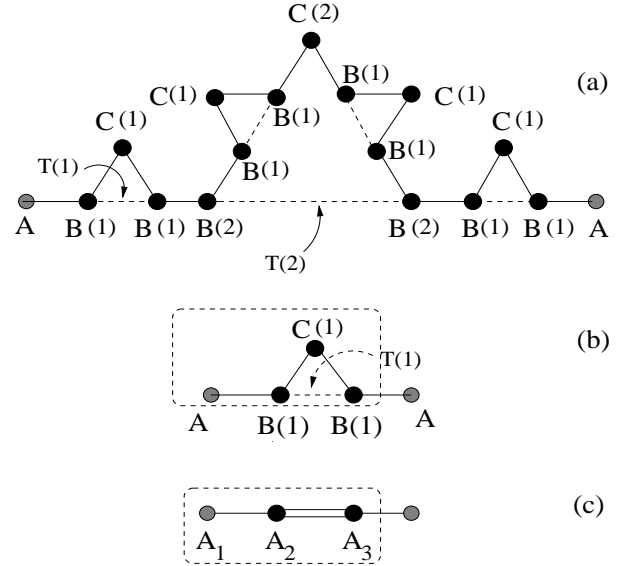}}\par}
\caption{ (a) A $2$nd generation Koch fractal with long range hopping. 
The sites $B(j)$ with $j=1$ and $2$ denote the `bulk sites' where $j$ 
denotes the stage of hierarchy. Accordingly, the longer ranged hopping 
integrals are assigned values $T(j)$, $j=1$,$2$. The boundary sites are marked as 
$A$ and the remaining sites are $C$. (b) A   
unit cell (inside the dashed box) of a periodic approximant built by repeating a
first generation Koch curve is shown. 
is enclosed in the dashed box, and, (c) The final {\it trimer} that is used for the trace 
analysis.}  
\label{lattice}
\end{figure}

Exceptional results however, started surfacing in the last decade of the twentieth 
century, when it was shown by Dunlap et al~\cite{dunlap} 
that a certain kind of {\it correlated disorder} 
could lead to a finite number of resonant eigenstates in an otherwise 
disordered one dimensional (1-d) chain. Such states are extended  
but, of non-Bloch character, and lead to a ballistic transport of electrons at special 
values of the energy. The so called {\it random dimer model} (RDM) proposed by 
Dunlap et al~\cite{dunlap} led to a thorough investigation of the effect of correlated 
disorder on the electronic and other excitonic states in one dimensional systems
~\cite{wu}-\cite{lyra}, where, 
ordinarily an un-correlated disorder renders all the states exponentially localized.

The effect of such short range correlations among the constituents in a one dimensional 
system has also led to exotic distribution of extended eigenstates in one dimensional 
quasi-periodic lattices~\cite{arun1}-\cite{macia2} where, owing to the 
self similarity of such systems, one 
encounters an infinity of such extended states in an infinite 1-d system. 

Deterministic fractals have also played their part. 
A subset of these structures has been shown to possess 
extended non-Bloch single particle states with a variety of transport 
properties~\cite{wang1}-\cite{schwalm4}. In contrast 
to the RDM or the quasi-periodic chains in 1-d, the fractal lattices do not exhibit any 
short range clustering of atomic sites (in the sense of the RDM) that could possibly 
lead to local {\it resonances} resulting in a de-localization of the wave function. Rather, 
the topology of the entire lattice plays the key role in giving rise to such extended 
eigenstates at special values of energy of an electron hopping around in 
these lattices.

The {\it extended} states in the cases referred to, appear, in general, in isolation 
in the energy spectrum where they coexist with an otherwise singular continuous part.   
However, in recent times extensive numerical studies have pointed out towards a possibility 
of having a continuum of such extended states in a class of deterministic 
fractals~\cite{arun3}-\cite{schwalm4}. 

A universal feature of such extended states, whether they arise in RDM models or in the 
fractals reported so far is that, their existence is not sensitive to the numerical values 
of the Hamiltonian parameters. The short range clustering 
of the atomic sites or the overall lattice topology determined the criteria of the 
existence of unscattered eigenstates. In the present communication we report an 
exceptional case where, well defined correlations between the values of a subset of the 
Hamiltonian parameters is {\it essential} for the construction of an {\it extended} eigenstate 
which has equal amplitude at every lattice point. We explain the phenomenon by solving the 
Schr\"{o}dinger equation on a Koch curve with hierarchical interaction~\cite{maritan}.
A Koch curve with a long range hierarchical distribution of electron-hopping models, 
in an approximate manner, a polymeric system~\cite{maritan}. We show that this model can lead to 
unusual spectral and electronic transport properties when the hierarchy parameter is allowed to vary,
mimicking (in a very crude sense though) {\it local distortions} in the system. 
The most astonishing result is that the same distribution of eigenstates, which makes the 
lattice conducting for a given set of the Hamiltonian parameters, represents a 
{\it completely localized} state with zero transmission across the Koch fractal.
In addition to this, we find that such contrasting results are obtained for an 
infinite variety of values of the hierarchy parameter. These values are distributed 
in a self-similar manner, as is apparent from a study of the two-terminal 
transport across a Koch fractal. We work within a tight binding formalism and employ 
a real space renormalization group (RSRG) decimation scheme to unravel these strange states at 
all scales of length, and to obtain the end-to-end transmission properties of a Koch fractal 
of any size.

In what follows, we explain our work. In section II the model and the method are described. 
Section III contains the results and the discussion, and in section IV we draw our conclusion.

\vskip .3in
\section{The model and the method}

We start by referring to Fig.~\ref{lattice}.  
\begin{figure}[ht]
\includegraphics[height=7cm,width=5cm,angle=-90]{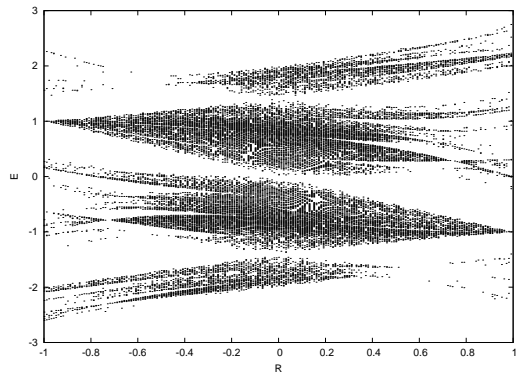}
\caption{Eigenvalue spectrum of a Koch fractal as a function of the 
hierarchical parameter $\lambda$ obtained from the trace of the 
transfer matrix for a $7$th generation fractal, taken as the 
`unit cell'.}  
\label{spectr}
\end{figure}

The tight binding Hamiltonian in the Wannier basis reads, 
\begin{equation}
H = \sum_{i}\epsilon_{i} c_{i}^{\dagger} c_{i}
+\sum_{ij} t_{ij} c_{i}^{\dagger} 
c_{j} 
+ h.c. 
\label{equ1}
\end{equation}
where, $\epsilon_{i}$ is the on-site energy of an electron at 
the site $i$. $t_{ij}$ is the 
hopping integral between neighboring sites. For nearest neighbor sites along the 
ridge of the Koch curve we assume $t_{ij} = t$, while the longer range hopping 
integrals can, in principle, take on any values. We shall illustrate our results 
for the specific case when the longer ranged hopping integrals assume values $T(n) = R^n t$, 
$R$ being the hierarchy parameter and $n$ represents the level of hierarchy.
For a Koch fractal of finite but 
arbitrarily large size, the on-site potential $\epsilon_i$ will be assigned values 
$\epsilon_{A}$ (the {\it extreme sites}) and  
$\epsilon_{B}(n)$, $\epsilon_{C}(n)$ etc for the bulk sites 
depending on their positions in the lattice as explained in Fig.~\ref{lattice}.

Owing to the self similarity of the lattice an application of the RSRG  
decimation method is used to scale the original system. An appropriate 
subset of sites are decimated leading to the 
recursion relations of the Hamiltonian parameters, viz,
\begin{eqnarray}
\epsilon_{A}' & = & \epsilon_{A} + \frac{(E - \xi) t^2}{\delta} \nonumber \\
\epsilon_{B}(n)' & = & \epsilon_{B}(n+1) +
2 \frac{(E - \xi) t^2}{\delta} \nonumber \\
\epsilon_{C}(n)' & = & \epsilon_{C}(n+1) + 2 \frac{(E - \xi) t^2}{\delta} \nonumber \\ 
t' & = & \frac{t^2 \tilde t}{\delta} \nonumber \\
T(n)'  & = & T(n+1)
\label{recur}
\end{eqnarray}
where, 
\begin{eqnarray}
\xi & = & \epsilon_{B}(1) + \frac{t^2}{E - \epsilon_{C}(1)} \nonumber \\ 
\tilde t & = & T(1) + \frac{t^2}{E - \epsilon_{C}(1)} \nonumber \\ 
\delta & = & (E - \xi)^2 - \tilde {t}^2
\end{eqnarray} 
The above set of recursion relations will be used to extract information about the eigenvalue spectrum 
and to discern the nature of the eigenstates in the light of the motivation presented before.

\begin{figure}[ht]
{\centering \resizebox*{11cm}{9cm}
{\includegraphics{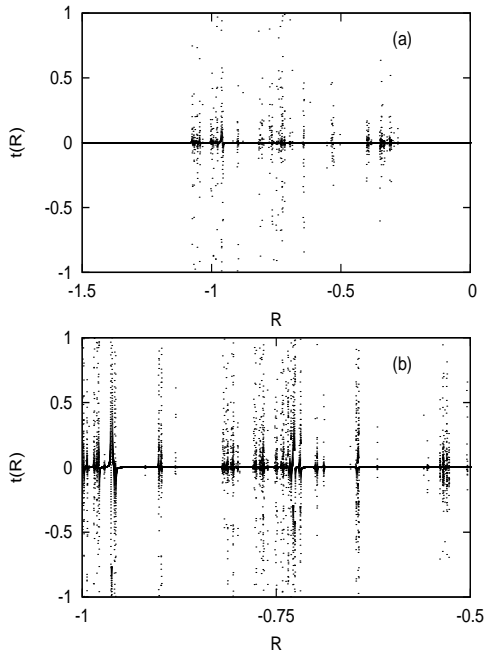}}\par}
\caption{(a) Renormalized hopping between the extremities of a $9$th generation 
Koch fractal against the hierarchy parameter $R$ at an energy $E = \epsilon_A + t$.
(b) A fine scan of the values of $R$ reveals the {\it fractal} nature of the 
distribution of the values of $R$.} 
\label{hop}
\end{figure}

\section{Results and discussion}
\subsection{The eigenvalue spectrum}

We try to gain some insight into the spectrum of an infinite Koch fractal in the 
presence of hierarchically distributed long range interactions. To achieve this, we 
proceed in the spirit of the one dimensional quasi-periodic lattices~\cite{koh}-\cite{luck}.
The basic idea is to repeat a finite generation Koch curve periodically, pick up the 
{\it unit cell} and extract the energy values for which the trace of the transfer matrix for the 
unit cell remains bounded by two~\cite{koh}. 
A prototype {\it unit cell} for a periodic approximant generated by repeating the 
first generation Koch curve is shown in Fig.~\ref{lattice}(b). To construct the transfer matrix and 
work out its trace in this case, 
the top vertex $C$ is decimated to reduce the system into a `triatomic molecule' $A_1$-$A_2$-$A_3$
(Fig.~\ref{lattice}(c))  
where,
\begin{eqnarray}
\epsilon_{A_1} & = & \epsilon_A \nonumber \\
\epsilon_{A_2} & = & \epsilon_B(1) + \frac{t^2}{E - \epsilon_C(1)} 
\end{eqnarray}
and, $\epsilon_{A_3} = \epsilon_{A_2}$. The `inter-atomic' hopping integrals are 
$t_{{A_1}-{A_2}} = t$, and $t_{{A_2}-{A_3}} = T(1) + t^2/[ E - \epsilon_C(1) ]$. 
The transfer matrix across this triatomic molecule is then easily calculated.
One can, in principle, begin with a fractal at any arbitrary generation, renormalize the 
parameters by sequentially applying the recursion relations Eq.~\ref{recur} $n$-times, bring it 
down to an effective first generation block and subsequently to an effective triatomic molecule 
as discussed above to get the trace of the transfer  
matrix. 
The spectrum of an infinite lattice is obtained when the {\it unit cell} itself 
becomes {\it infinitely large}.

We present the result in Fig.~\ref{spectr} where the 
{\it unit cell} is a seventh generation Koch fractal with hierarchically distributed values of 
the long range hopping integrals $T(n) = R^n t$. The `band structure' is displayed as the hierarchy 
parameter $R$ is varied. The behavior seems to saturate, at least becomes indistinguishable, 
for unit cells of larger size. So, the present figure can be taken to represent a very good 
approximation to the true eigenvalue spectrum in the infinite lattice limit.

With low absolute values of the hierarchy parameter $R$, the spectrum exhibits clustering of 
eigenvalues, giving rise to the formation of sub-bands. The sub-bands are separated by gaps 
which widen as $R$ increases. However, for $R$ quite large compared to the nearest neighbor 
hopping $t$, the sub-bands shrink in their widths and global gaps open up. 
An indefinite increase in the value of $R$ finds the spectrum consisting of a sparse 
distribution of points.
\begin{figure}[ht]
{\centering \resizebox*{11cm}{9cm}
{\includegraphics{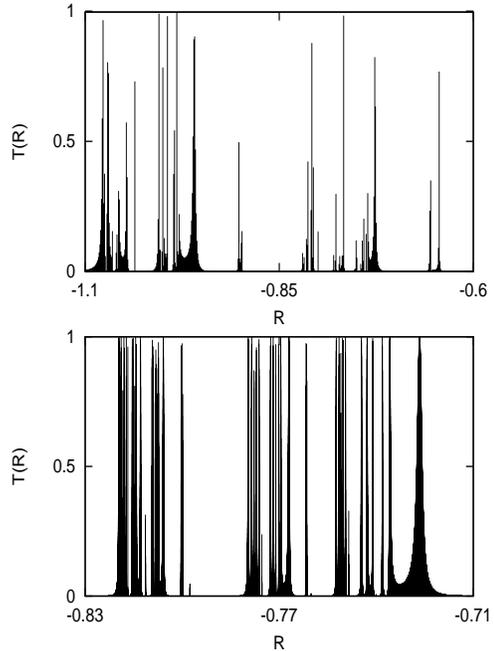}}\par}
\caption{(a) The transmission coefficient of a $7$th generation Koch fractal 
with hierarchical interactions $-1 \le R \le -0.6$, and (b) a finer scan 
for $-0.83 \le R \le -0.71$. The fractal distribution of $T$ against $R$ is to be 
noted.}  
\label{trans1}
\end{figure}
\begin{figure}[ht]
{\centering \resizebox*{11cm}{9cm}
{\includegraphics{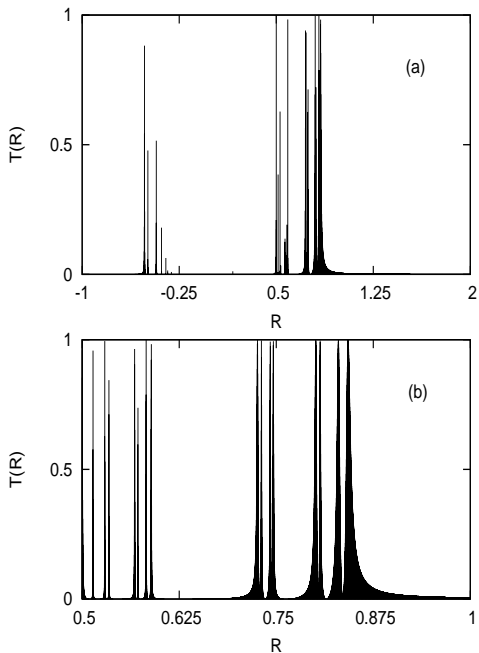}}\par}
\caption{Transmission coefficient of a $6$th generation 
Koch fractal against the hierarchy parameter $R$. Here, the energy 
has been obtained from the equation $E = \epsilon_{A}' + t'$, that is, 
from a one step renormalized lattice. The variation with $R$ and its fine scan 
revealing again a self similar character, are shown in figures (a) and (b) 
respectively.} 
\label{trans2}
\end{figure}

\subsection{The atypically {\it extended} eigenstates}

In this discussion we highlight certain aspects of a subset of eigenstates 
on a Koch fractal of finite but arbitrarily large generation. Let us choose 
the energy of the electron (the Fermi energy) as $E = \epsilon_A + t$. With this choice, 
it is easy to verify that one can construct an 
eigenstate whose amplitude at each site of the Koch fractal 
will be identical (normalized to unity, say) if we select 
\begin{eqnarray}
\epsilon_C(n) & = & \epsilon_A - t \nonumber \\
\epsilon_B(n) + T(n) & = & \epsilon_A - t 
\label{equ2}
\end{eqnarray}
for all $n \geq 1$. The above choices of the on-site potentials and the hierarchical 
hopping integrals $T(n)$ define a correlation between the values of the parameters of 
the Hamiltonian. Interestingly, we can see that once the on-site potential $\epsilon_A$ 
of the extreme sites and the nearest neighbor hopping matrix element $t$ are fixed, 
$\epsilon_B(n)$ and $T(n)$ can, in principle, be chosen from a set of random values, but in 
such a way, so as to satisfy the desired correlation, viz, $\epsilon_B(n) + T(n)  =  \epsilon_A - t$.
Since, with this choice, the probability of finding the electron is same on every lattice point, 
it is not unusual to assume these states to be {\it extended} in the conventional language. 
However, a study of the flow of the nearest neighbor hopping integral $t$ under successive RSRG steps 
reveals a deeper story. It is known that, if, corresponding to certain value of the energy $E$, 
the hopping integral remains non-zero under RSRG iteration, it implies that the nearest neighbors 
are `connected' at all scales of length. Hence, the corresponding eigenstate can legitimately be 
designated as an {\it extended} state. In the present case, the result of this investigation is 
found to be strongly sensitive to the initial choice of the parameters $\epsilon_B(n)$ and 
$T(n)$. 

To clarify the above idea, we consider the particular situation where the longer range hopping integrals 
$T(n)$ are distributed in a hierarchical fashion, viz, $T(n) = R^n t$. We set $E = \epsilon_A + t$,
and take $\epsilon_C(n) = \epsilon_A - t$ for all $n$. The choice of $\epsilon_B(n)$ is then 
restricted by the desired correlation as it appears in the last of the equations in Eq.~\ref{equ2}.
A Koch fractal, on successive renormalization by an appropriate number of steps may be 
reduced to an effective diatomic molecule comprising of the end sites only. The effective 
hopping between the surviving end sites will decide on the {\it extendedness} of the wave function 
at that particular energy. Most interestingly we find that, the flow of the hopping 
integral $t$ under successive iterations is strongly sensitive to the choice of the hierarchy 
parameter $R$ (which automatically fixes $\epsilon_B(n)$). We present one such result in 
Fig.~\ref{hop}, where the end-to-end renormalized hopping in a $9$th generation Koch fractal 
is plotted as a function of the hierarchy parameter $R$. We have taken  $E = 1$ and $T(n) = R^n t$ 
with $\epsilon_A = 0$, $t = 1$, $\epsilon_C(n) = -1$ (for all $n$), and 
$\epsilon_B(n) = \epsilon_A - t - T(n)$. It must be appreciated that for every value of $R$, with 
the above conditions being satisfied, the wave function displays an identical pattern of 
amplitudes on all sites. yet, we find that the end-to-end hopping becomes zero or remains finite 
depending on the values of $R$ (and hence the values of $\epsilon_B(n)$). In addition to that, 
a finer scan of any of the intervals of $R$ reveals that the values of the hierarchy parameter 
for which the end-to-end hopping across a Koch fractal of any generation remains non-zero 
(a true extended state), are distributed in a self-similar, fractal pattern. Every value of 
$R$, defines a new point in the parameter space $[\epsilon_A, \epsilon_B(n), \epsilon_C(n), T(n), t]$. 
therefore, for $E = \epsilon_A + t$, we come across a {\it fractal} distribution 
of points in the parameter space.
The features described above are reflected in the end-to-end transmission across a finite 
Koch curve with hierarchical interactions. The results are presented below.

\subsection{The transmission coefficient}

We have examined the two-terminal transport of a Koch fractal of an arbitrary 
generation by placing the lattice between two semi-infinite 
perfectly periodic leads characterized by a constant on-site potential 
$\epsilon_0$ and nearest neighbor hopping integral $t_0$. A $k$-th generation 
fractal is reduced, by applying the decimation recursion 
relations Eq.~\ref{recur} $k$ times, to an effective diatomic molecule 
comprising of the extreme sites $A-A$ with renormalized on-site potential $\epsilon_A^{(k)}$  
and renormalized `inter atomic' hopping integral $t^{(k)}$. The transmission coefficient 
$\tau$ is then 
obtained from the well known formula~\cite{stone}, 
\begin{widetext}
\begin{equation}
\tau = \frac{4 \sin^2 qa}{[ M_{12} - M_{21} + (M_{11} - M_{22}) \cos qa ]^2 +
 [ M_{11} + M_{22} ]^2 \sin^2 qa}
\label{tranform}
\end{equation}   
\end{widetext}
Here, 
\begin{eqnarray}
M_{11} & = & \frac{[ E - \epsilon_A^{(k)} ]^2}{t_0 t^{(k)}} - \frac{t^{(k)}}{t_0} \nonumber \\
M_{12} & = & - \frac{E - \epsilon_A^{(k)}}{t^{(k)}} \nonumber \\
M_{21} & = & - M_{12} \nonumber \\
M_{22} & = & - \frac{t_0}{t^{(k)}} 
\end{eqnarray}
The lattice spacing $a$ in the leads is taken to be unity, and $q = \cos^{-1}[(E-\epsilon_0)/(2t_0)]$.
Throughout the calculations we take $\epsilon_0 = 0$ and $t_0 = 1$.

Fig.~\ref{trans1} (a) displays the transmission across a $7$th generation 
fractal for $E = \epsilon_A + t$ for a certain choice of the range of the 
hierarchy parameter $R$. For every value of $R$ in this range, we have 
taken $\epsilon_B(n) = \epsilon_A - t - T(n)$ which makes the amplitudes of the 
wave function same at every site (normalized to one). The spectrum consists of 
gaps and resonant transmissions indicating that, for certain values of $R$ the 
$7$th generation Koch network may become completely transparent to an incoming 
electron with the above energy, while that same electron finds the network totally 
opaque for other values of $R$. A fine scan of a sub-interval of $R$ reveals that 
the values of $R$ presenting the above feature are indeed distributed in a 
self-similar, fractal manner. This is illustrated in 
Fig.~\ref{trans1}(b) and is in agreement with the initial 
observations made based on the recursive behavior of the nearest neighbor hopping.

One can construct eigenstates for which the amplitudes will be same (unity) at 
all the sites of a renormalized Koch fractal as well. The procedure outlined above for the 
lattice at the bare length scale is to be implemented on a renormalized version of the 
system. For example, we can set $E = \epsilon_A' + t'$ on a one step renormalized 
lattice.  
The eigenvalues of the desired 
states are then extracted by solving the polynomial equation 
\begin{equation}
E - F \left( \epsilon_A, t, \epsilon_B(1), \epsilon_C(1), T(1) \right ) = 0
\label{roots}
\end{equation}
where, 
\begin{widetext}
\begin{equation}
F = \frac{\epsilon_A E^2 + E \left [ t^2 - \epsilon_A \left ( \epsilon_B(1) + 
\epsilon_C(1) + T(1) \right ) \right ] + \epsilon_A \left [ \epsilon_B(1) \epsilon_C(1) - 2 t^2 +
\epsilon_C(1) T(1) \right ] - t^2 \epsilon_C(1)}{E^2 - 
\left [ \epsilon_B(1) + \epsilon_C(1) + T(1) \right ] E + \epsilon_B(1) \epsilon_C(1) - 2 t^2 +
\epsilon_C(1) T(1)}
\label{fac}
\end{equation}
\end{widetext}
The potentials and the hopping integrals appearing as the arguments of $F$ can 
be chosen independently. 
It is to be appreciated that the roots of the equations are free from $\epsilon_B(n)$ and 
$\epsilon_C(n)$ and $T(n)$ for $n \geq 2$. 
We shall now demand $\epsilon_C(1)' = \epsilon_A' - t'$ which leads to an equation 
defining $\epsilon_C(2)$ in the original lattice, viz, 
\begin{equation}
\epsilon_C(2) = \left [ \epsilon_A' - t' - \Delta \right ]_{E = \epsilon_A'+t'}
\label{c2}
\end{equation}
where, 
\begin{widetext}
\begin{equation}
\Delta = \frac{t^2}{E + T(1) - \epsilon_B(1)} + \frac{[E - \epsilon_C(1)] t^2}
{E^2 -E [\epsilon_C(1) + T(1)] + \epsilon_C(1) [ \epsilon_B(1) + T(1) ] -2 t^2}
\label{delta}
\end{equation}
\end{widetext}
The energy $E$ is chosen from the solutions of Eq.~\ref{roots}.
This settles the initial value of $\epsilon_C(2)$ we should begin with. 
To mimic the 
earlier situation in the original scheme (the bare 
length scale) we now set 
$\epsilon_C(n) = \epsilon_C(2)$, 
as obtained above, for all $n \geq 2$ in the original Koch geometry.. 

In addition to this,  we should set 
$\epsilon_B(n)' + T(n)' = \epsilon_A' - t'$ for $n \ge 1$, which leads to the selection of the 
un-primed values of $\epsilon_B(n)$ and $T(n)$ such that, 
$\epsilon_B(n) + T(n) = ( \epsilon_A' - t' - \Delta )_{E = \epsilon_A'+t'} = \epsilon_C(2)$, 
for $n \geq 2$ on the original (bare length scale) lattice.
The entire scheme now works just as it worked in the original scale of length. The process can, 
in principle, go on and every time we extract the energy eigenvalue by solving the 
polynomial equation $E - \epsilon_A^{(k)} - t^{(k)} = 0$ from the $k-$th stage of renormalization, 
we need to pre-define  a bigger parameter space to begin with. The energy of course remains 
independent of $\epsilon_B(n)$, $\epsilon_C(n)$ or $T(n)$ for $n > k$. The scheme works, the 
correlations become much more complicated though.

\section{Conclusion}
In conclusion, we have unraveled a set of atypical extended states in Koch fractals 
possessing hierarchically long range interactions. The states have identical amplitudes at 
all atomic sites in the lattice for mutually correlated values of a certain set of the  
Hamiltonian parameters. The same distribution of amplitudes on a Koch curve of arbitrary 
generation may represent either a completely transparent state or a strictly localized one 
depending on the choice of the parameters. This fact has been illustrated by choosing 
the long range hopping integrals to be distributed in a hierarchical, but 
deterministic fashion, and by tuning the 
hierarchical parameter appropriately and according to the terms dictated by the correlation.
\vskip .3in
\begin{center}
{\bf ACKNOWLEDGMENT}
\end{center}
\vskip .25in
I thank Enrique Maci\'{a} for an early discussion on this work, and Bibhas 
Bhattacharyya for helpful suggestions.
                                                                            
\end{document}